\documentclass[a4paper,fleqn,usenatbib]{mnras}

\usepackage{newtxtext,newtxmath}

\usepackage[T1]{fontenc}
\usepackage{ae,aecompl}

\usepackage{graphicx}	
\usepackage{amsmath}	
\usepackage{amssymb}	

\newcommand{\snr}{{SNR G15.9+0.2}}
\newcommand{\xmm}{{\it XMM-Newton}}

\newcommand{\chandra}{{\it Chandra}}
\newcommand{\herschel}{{\it Herschel}}
\newcommand{\spitzer}{{\it Spitzer}}

\newcommand{\nh}{\mbox {$N_{\rm H}$}}
\def\la{\mathrel{\hbox{\rlap{\hbox{\lower4pt\hbox{$\sim$}}}\hbox{$<$}}}}

\title[\snr]{Infrared and X-ray study of the Galactic \snr}

\author[M. Sasaki et al.]{
Manami Sasaki,$^{1}$\thanks{E-mail: manami.sasaki@fau.de}
Minja M. M\"akel\"a,$^{1}$
Dmitry Klochkov,$^{2}$
Andrea Santangelo,$^{2}$
\newauthor and Valery Suleimanov$^{2,3}$
\\
\\
$^{1}$Dr. Karl Remeis-Sternwarte,
Erlangen Centre for Astroparticle Physics,
Friedrich-Alexander-Universit\"at Erlangen-N\"urnberg,
Sternwartstra{\ss}e 7, \\
D-96049 Bamberg, Germany\\
$^{2}$
Institut f\"ur Astronomie und Astrophysik,
Universit\"at T\"ubingen,
Sand 1,
D-72076 T\"ubingen, Germany,\\
$^{3}$
Kazan (Volga region) Federal University, 
Kremlevskaya street 18, 
Kazan 420008, Russia
}

\date{Accepted XXX. Received YYY; in original form ZZZ}

\pubyear{2016}

\begin{document}
\label{firstpage}
\pagerange{\pageref{firstpage}--\pageref{lastpage}}
\maketitle

\begin{abstract}
G15.9+0.2 is a Galactic shell-type supernova remnant (SNR), which was detected 
in radio and has been confirmed in X-rays based on \chandra\ observations. An 
X-ray point source CXOUJ181852.0-150213 has been detected and suggested to be 
an associated neutron star. In a recent study, we have confirmed the source to 
be a central compact object (CCO). 
We have studied the SNR using high-resolution X-ray data taken with \chandra\
in combination with infrared (IR) data in order to understand its emission and 
to derive its physical parameters.
This will also help to constrain, e.g., the age of the CCO and the environment 
in which it was born.
The spectral analysis of the X-ray emission using the new \chandra\ data and 
the comparison to the IR data have shown that the SNR is 
relatively young with an age of a few thousand years and that its emission is 
dominated by that of shocked interstellar medium (ISM).
However, the analysis of the spectrum of the bright eastern shell shows that
there is an additional emission component with enhanced abundances of 
$\alpha$ elements and Fe, suggesting ejecta emission.
The multi-wavelength emission is consistent with 
\snr\ expanding in an ISM with a density gradient, while there is also colder 
material located in front of the SNR, which absorbs its thermal X-ray emission 
in the softer bands.
\end{abstract}




\section{Introduction}\label{intro}

Massive stars inject matter and energy into the ambient interstellar
medium (ISM) through their stellar winds and, at the end of their lives, 
in supernova explosions. The cores of massive stars collapse and form
either a neutron star, or in fewer cases, a black hole. 
The spherically expanding shock waves heat and ionise the ambient ISM and 
the stellar ejecta and form supernova remnants (SNRs).  These extended 
objects can emit radiation over the entire observable electromagnetic 
spectrum, depending on the physical processes that play the dominant 
role, e.g., heating by shock waves, distribution of stellar ejecta, 
evaporation of denser regions of the ISM, or particle acceleration in 
the shock waves. 
Shocks are in general caused by an `external' impact in a compressible medium,
like a supernova explosion. In an ionised, low-density medium, there will be
no collisions between the particles
within the width of the shock front,
the shock is rather mediated
by interactions of charged particles and electromagnetic fields. 
The heated ions radiate copious X-ray emission, while
electrons accelerated in the shocks and interacting with the compressed
magnetic fields emit radio synchrotron emission.
The shocked plasma will heat the cold interstellar dust through collisions,
which will emit in the infrared (IR).
Therefore, combining X-ray observations with lower energy data will help us
to understand the physics of SNR shocks and the properties of the ISM
in which the SNR is expanding.

The Galactic \snr\ was first detected in radio and classified as an SNR by
\citet{1975AuJPA..37....1C}. The flux density at 6 cm is $1.95 \pm 0.29$ Jy
with an index of $\alpha = -0.63 \pm 0.03$ 
\citep[][and references therein]{2011A&A...536A..83S}. 
Several radio observations had been carried out, but it
was not until 2005 that X-ray 
observations with the \chandra\ X-ray 
observatory were performed and a shell-type SNR was identified in X-rays 
\citep{2006ApJ...652L..45R}. The X-ray spectrum is indicative of thermal
emission in non-equilibrium ionisation with higher abundance for sulfur.
The spectrum is highly absorbed with a foreground H column density of 
\nh\ = $4 \times 10^{22}$ cm$^{-2}$. In addition, the high-resolution \chandra\
data revealed a point source which was suggested to be the neutron star formed
in the supernova explosion.

We carried out an additional \chandra\ observation in 2015 for 90 ks to 
study the point source CXOUJ181852.0--150213 and the SNR shell
\citep{2016A&A...592L..12K}. Based on the
new data we showed that CXOUJ181852.0--150213 is a neutron star and belongs
to the class of central compact objects (CCOs), which are believed to be 
weakly magnetised
cooling neutron stars with atmospheres emitting thermal X-rays. 
So far, dust emission has been observed only from a few SNRs with a CCO
\citep[e.g.,][]{2016MNRAS.458.2565D,2017MNRAS.465.3309D}.
In this work, we present the study of the X-ray and IR emission of \snr.

\section{Data}

\subsection{X-rays}

\begin{table}
\centering
\caption{
\label{obstab}
List of \chandra\ observations of \snr.}
\centering
\begin{tabular}{rllrrl}
\hline\hline 
ObsID & Instrument & Mode & Exposure & Year & PI \\
 & & & [ks] & & \\
\hline
5530 & ACIS-S & VFAINT & 9 & 2005 & Reynolds \\
6288 & ACIS-S & VFAINT & 5 & 2005 & Reynolds \\
6289 & ACIS-S & VFAINT & 15 & 2005 & Reynolds \\
16766 & ACIS-S & VFAINT & 92 & 2015 & Klochkov \\
\hline\hline 
\end{tabular}
\end{table}

\begin{figure*}
\centering
\includegraphics[trim=13 13 13 13,clip,width=\textwidth]{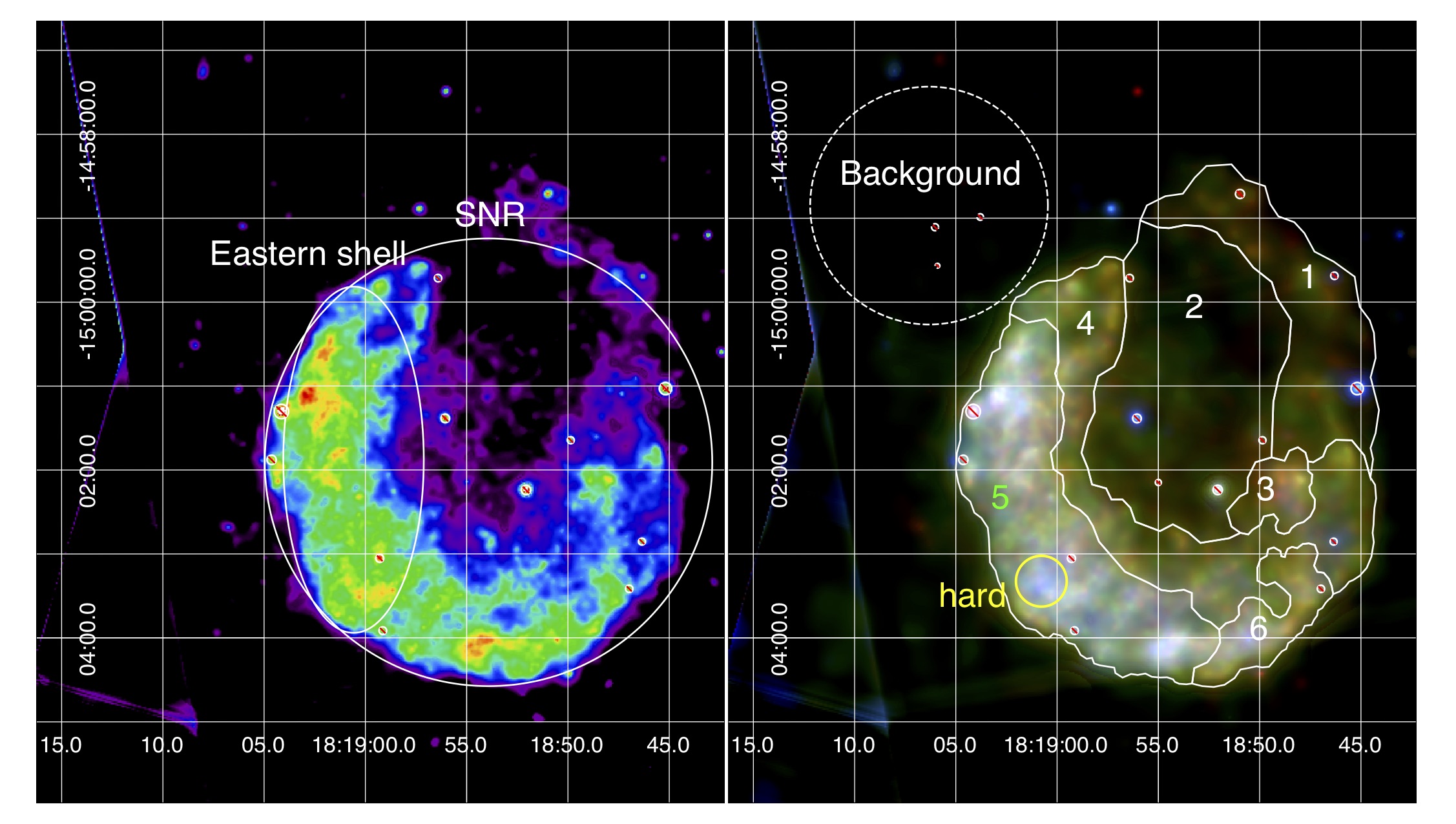}
\caption{
Broad-band mosaic X-ray image (0.3 -- 8.0 keV)
of \snr\ taken with \chandra\ ACIS-S (left)
and three-colour image 
(red  = 1.0 -- 2.0 keV, green = 2.0 -- 3.5 keV, and 
blue = 3.5 -- 7.0 keV, right).
Regions used for the extraction of spectra are also shown.
Point sources, visible in these images, were removed when the spectra were
extracted.
The images are all shown in equatorial coordinates. North is up, west is to 
the right.
\label{images1}}
\end{figure*}

The \snr\ was observed in 2005 with \chandra\ for a total of $\sim$30 ks (see 
Table \ref{obstab}). The analysis of the data and the results have been 
published by \citet{2006ApJ...652L..45R}.
In 2015 we obtained new proprietary data from an observation with an
exposure of 92 ks (PI: D. Klochkov). 
\snr\ was observed on the ACIS-S3 chip in all observations.
We have reprocessed the data of the observations 5530, 6288, 6289, and 
16766 using CIAO version 4.8 and CALDB version 4.7.1. The study of the main 
target of the new observation, which was the central compact object 
CXOUJ181852.0--150213, has been published by \citet{2016A&A...592L..12K}.
We combined all data to create exposure-corrected images of the SNR 
in the broad band (0.3 -- 8.0 keV) and in sub-bands (0.3 -- 1.0 keV, 1.0 --
2.0 keV, 2.0 -- 3.5 keV, and 3.5 -- 7.0 keV). 
As has already been shown by \citet{2006ApJ...652L..45R}, \snr\ is heavily
absorbed  (\nh $> 10^{22}$ cm$^{-2}$). The mosaic image of a total of
120 ks confirms that there is no significant emission below
1 keV. In Figure \ref{images1}, 
an exposure-corrected image in the band of
0.3 -- 8.0 keV and a three-colour image with red = 1.0 -- 2.0 keV, 
green = 2.0 -- 3.5 keV, and blue = 3.5 -- 7.0 keV are shown.
The X-ray shell is bright in the east and the south, while there is only
faint filamentary emission in the northwest. If one draws a circle around
the SNR to find the extent under the assumption that the expansion was 
homogeneous in all direction, the northern filaments are located well outside 
the circle. The faintness of the northern X-ray emission and its larger 
distance from the center of the SNR suggest that the ISM density
is lower in the north and the shock is expanding faster in this direction.

\subsection{H$\alpha$ image}

To study the distribution of the colder ISM at the position of the SNR and
verify whether the SNR has developed radiative filaments, we
obtained H$\alpha$ and short red 
continuum images of the 
SuperCOSMOS H$\alpha$ survey \citep[SHS,][]{2005MNRAS.362..689P}, 
carried out by the Wide Field Astronomy Unit of the 
Institute for Astronomy of the Royal Observatory, Edinburgh. 
The continuum-subtracted H$\alpha$ image is shown in Fig.\,\ref{images2} 
(top left) with contours from the \chandra\ mosaic broad band image.
No obvious H$\alpha$ emission related to the SNR can be found, which would 
have been a sign for the SNR being radiative. The image rather
shows the distribution of warm gas on the line of sight. 
The H$\alpha$ emission is less bright on the eastern side (on 
the left-hand side in Fig.\,\ref{images2}), indicating that material 
located on the line of sight is absorbing the background emission. 
As will be shown in Sect.\,\ref{xspectra}, the X-ray emission of the
SNR indicates higher absorption in this region, confirming that
there is material located in front of the SNR.

\subsection{Infrared}

The \snr\ has also been detected in the IR and shows a clearly
shell-like structure similar to its morphology in X-rays. 
We have obtained the \spitzer\ InfraRed Array Camera 
\citep[IRAC;][]{fazioetal04} and Multiband Imaging Photometer 
\citep[MIPS;][]{riekeetal04} images from the \spitzer\ Heritage Archive. 
The IRAC images taken in October 2004 (program ID: 146) are a part of the GLIMPSE 
legacy survey \citep{benjaminetal03, churchwelletal09}, and the MIPS images 
were taken in September 2005 (program ID: 20597) as a part of the MIPSGAL 
survey \citep{careyetal09}. The pipeline-processed post-BCD images have been 
calibrated and mosaicked using pipelines S18.25 and S18.12 for IRAC and MIPS, 
respectively, and they are in units MJy/sr. The estimated flux calibration 
uncertainty for the pipeline products is 4~\%\footnote{MIPS instrument 
handbook, https://irsa.ipac.caltech.edu/data/SPITZER/ docs/mips/mipsinstrumenthandbook/}.

\begin{figure*}
\centering
\includegraphics[trim=50 50 50 50,clip,width=.47\textwidth]{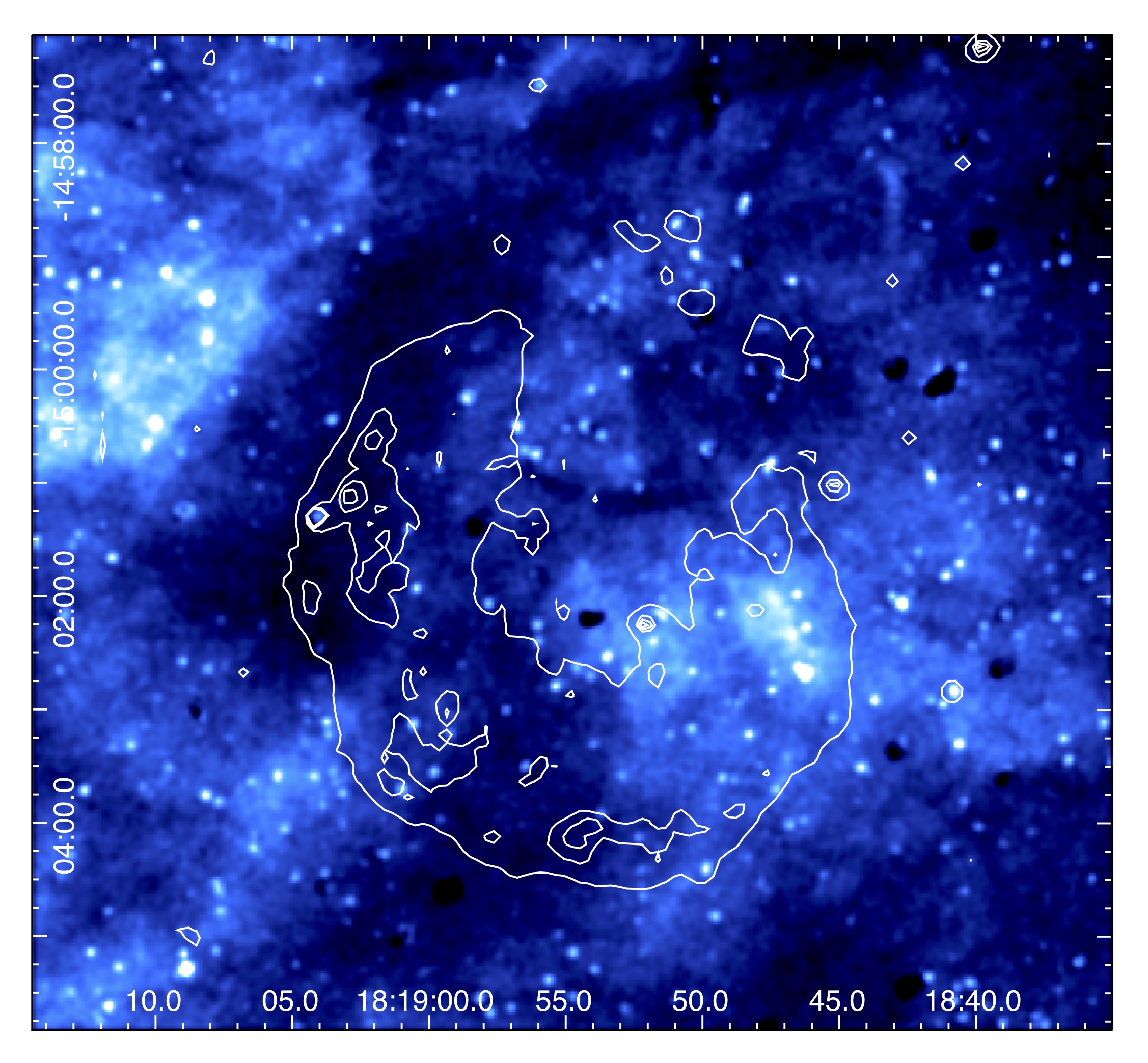}
\includegraphics[trim=50 50 50 50,clip,width=.47\textwidth]{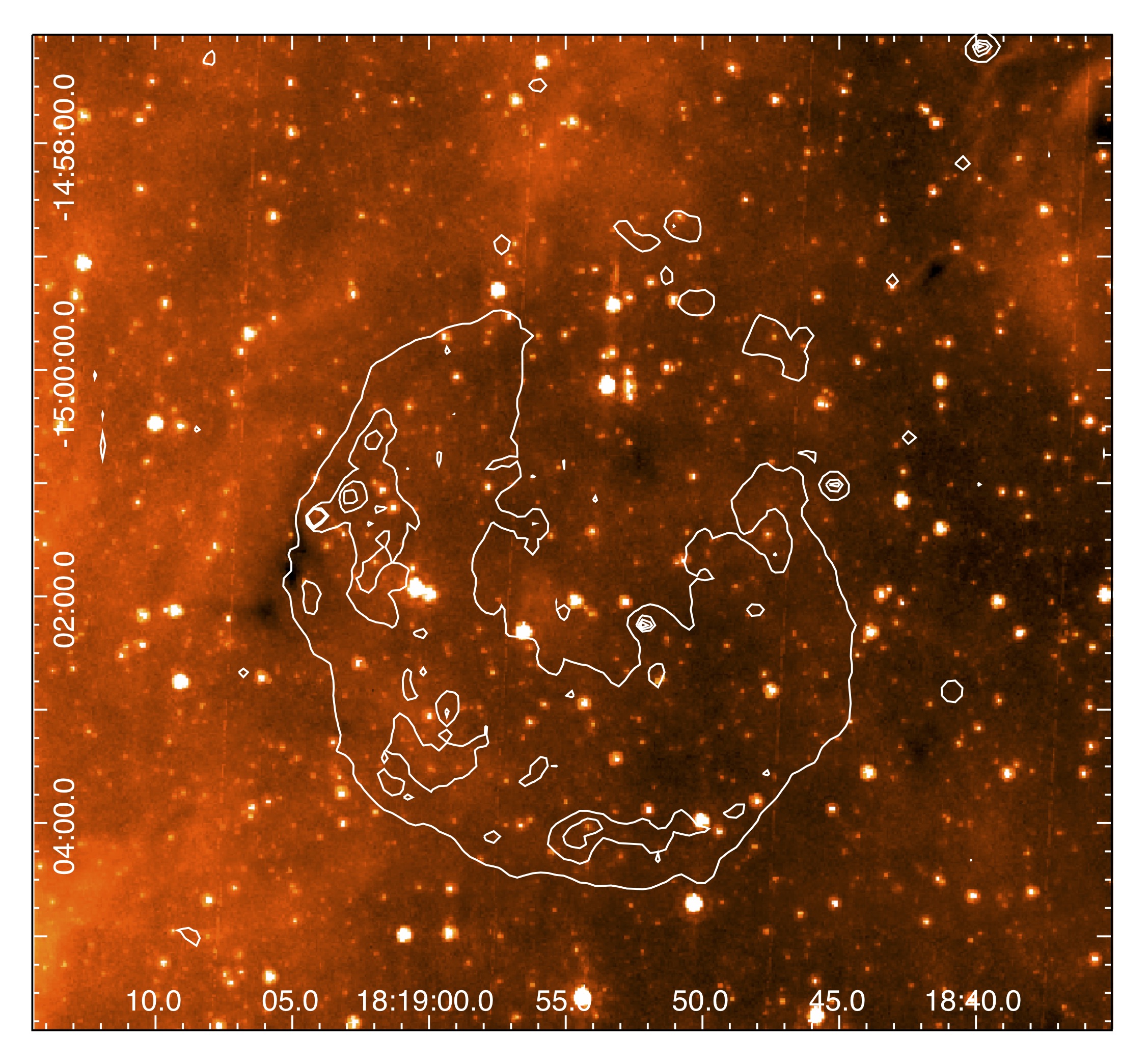}\\
\includegraphics[trim=50 50 50 50,clip,width=.47\textwidth]{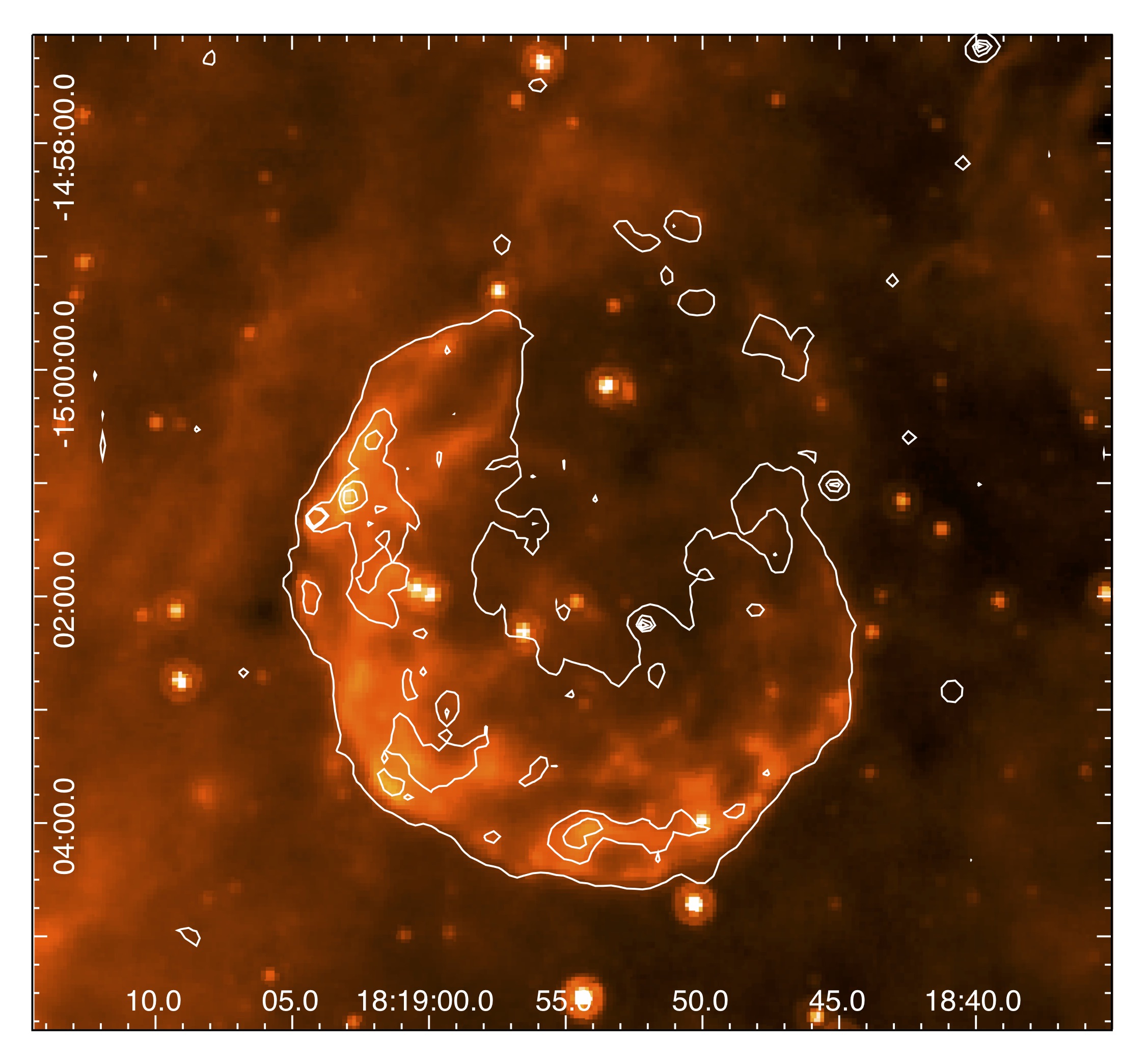}
\includegraphics[trim=50 50 50 50,clip,width=.47\textwidth]{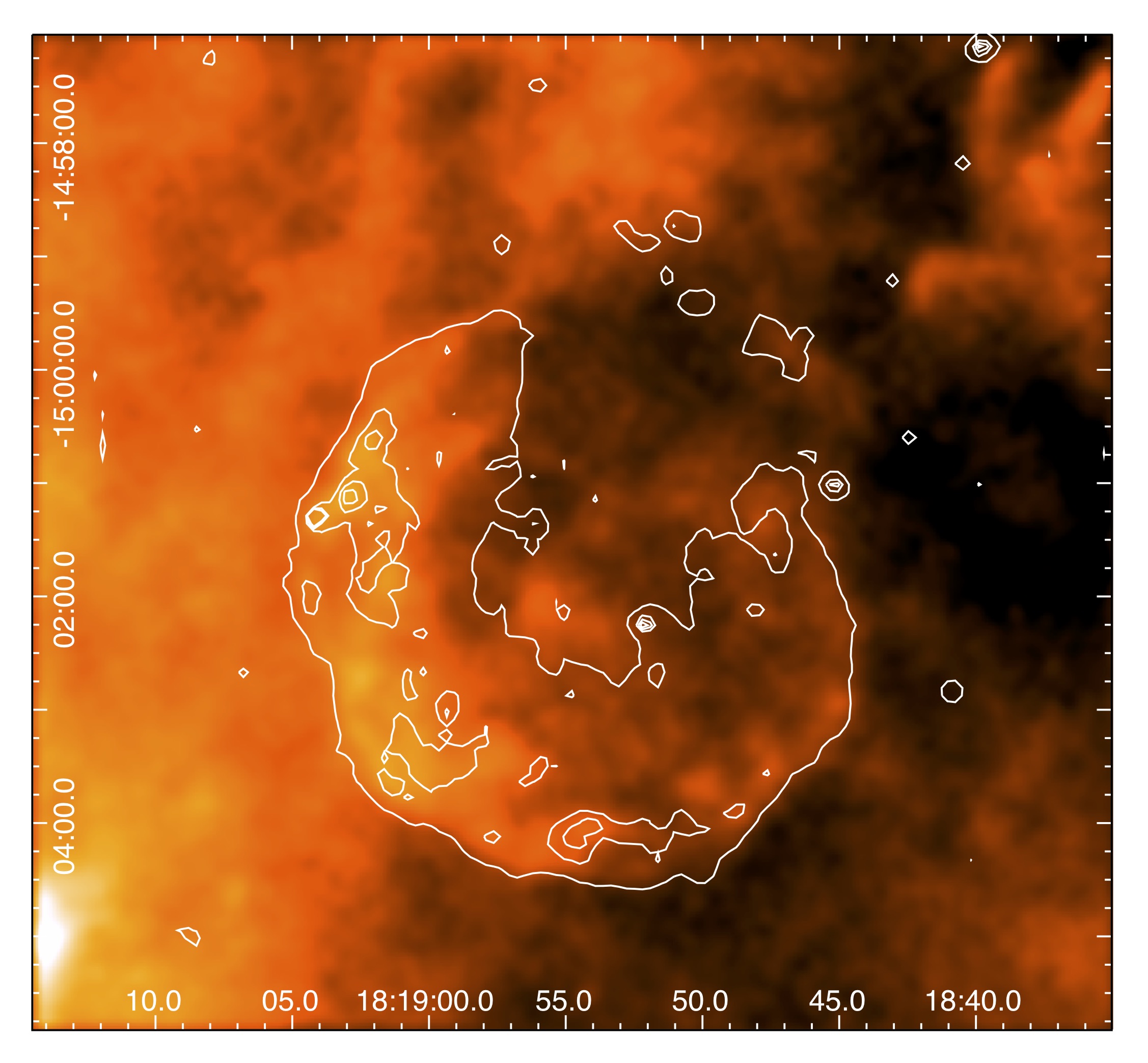}\\
\includegraphics[trim=42 50 57 50,clip,width=.47\textwidth]{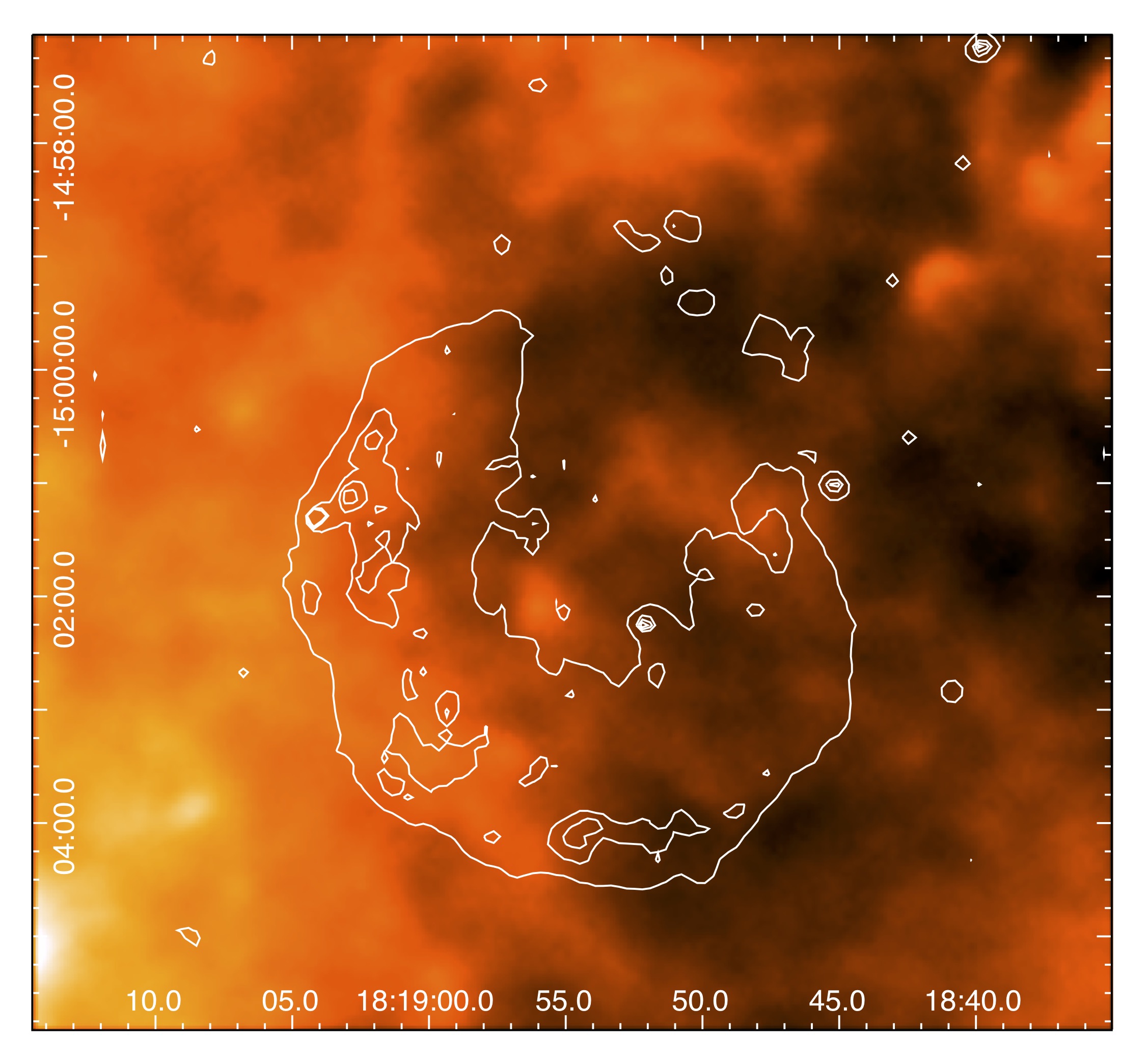}
\includegraphics[trim=42 50 57 50,clip,width=.47\textwidth]{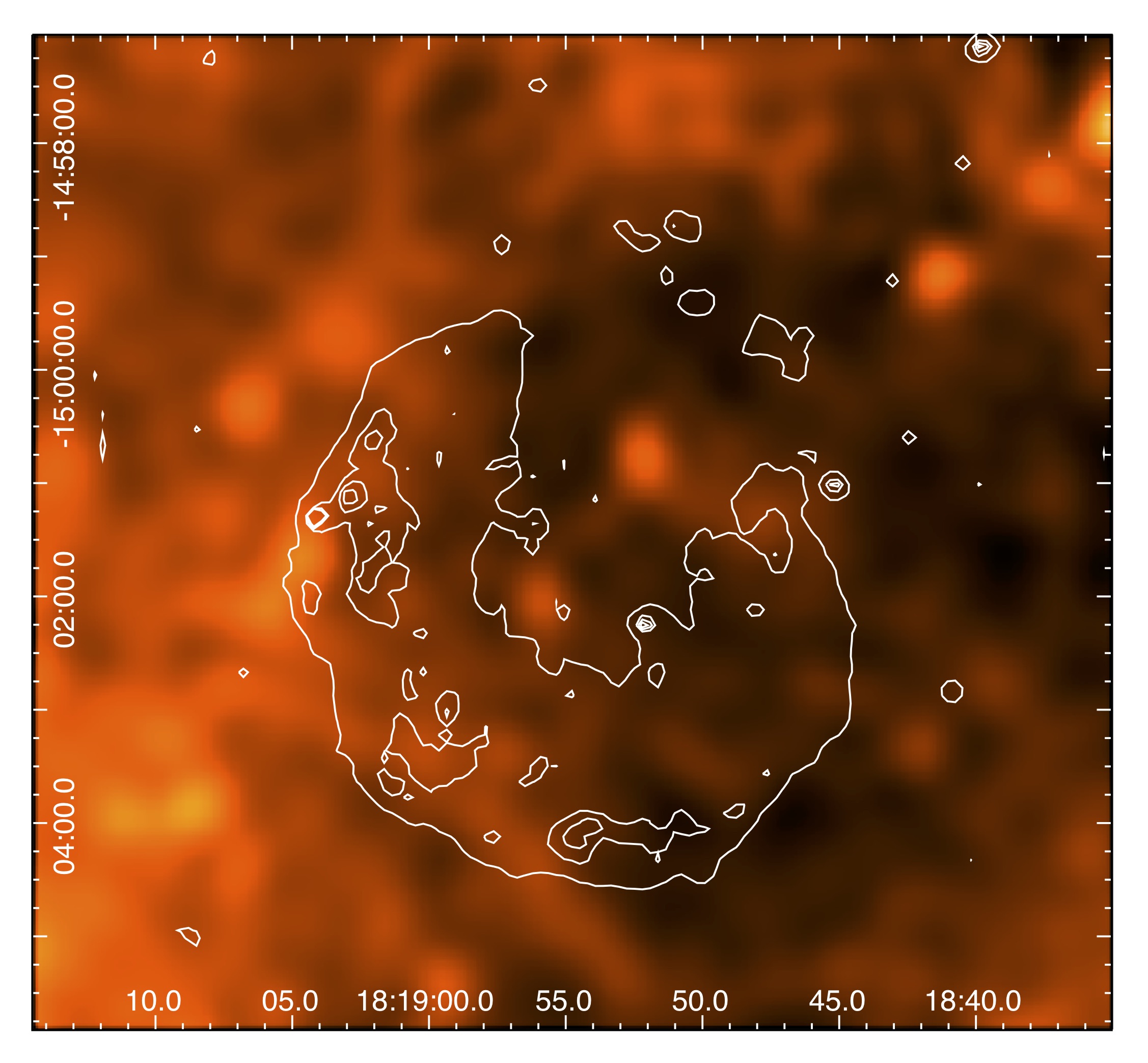}
\caption{
{\it Top}: SHS H$\alpha$ image (continuum-subtracted, left)
and \spitzer\ IRAC 8.0~$\mu$m image (right).
{\it Middle}:
\spitzer\ MIPS 24~$\mu$m (left) and \herschel\ PACS 
70~$\mu$m images (right), in which the SNR is well visible. 
{\it Bottom:} \herschel\ PACS 160~$\mu$m (left) and SPIRE 350~$\mu$m images
(right).
All images are presented in linear scale. The white contours are
\chandra\ count rates from 2$\times 10^{-8}$ cts s$^{-1}$ to 
2$\times 10^{-7}$ cts s$^{-1}$ in steps of 6$\times 10^{-8}$ cts s$^{-1}$.
\label{images2}}
\end{figure*}

We also extracted level 2.5 \herschel\ images taken with the Photodetector 
Array Camera \citep[PACS;][]{poglitschetal10} and with the Spectral and 
Photometric Imaging Receiver \citep[SPIRE;][]{griffinetal10} from the 
\herschel\ Science Archive (OBSIDs 1342218997, 1342218998). The PACS images 
were taken at bands 70 and 160~$\mu$m and the SPIRE images at bands 250, 350, 
and 500~$\mu$m. The PACS 160 and SPIRE 350~$\mu$m images are shown in
Fig. \ref{images2} (bottom). 
The observations were performed in April 2011 
as a part of the Hi-GAL survey \citep{molinarietal10}. 
The \herschel\ 70~$\mu$m images were used in this work instead of the 
\spitzer\ 70~$\mu$m images because the spatial resolution of the PACS images 
is higher (5\farcs2 vs. 16\arcsec). The PACS images are in Jy/px format, and 
the extracted SPIRE images have been HiRes-processed \citep{xuetal14}. The 
absolute flux calibration error in PACS images is estimated to be 
5~\%\footnote{PACS instrument and calibration web pages, http://herschel.esac.esa.int/ twiki/bin/view/Public/PacsCalibrationWeb}.

\snr\ is very clearly seen at 24~$\mu$m and 70~$\mu$m with a shell-like
structure similar to the X-ray emission.
The \spitzer\ MIPS 24~$\mu$m image shows the structure of the SNR best
(see Fig.\,\ref{images2}, middle left).
At 24~$\mu$m the southern and eastern parts of the shell are bright, while at
70~$\mu$m the eastern side is bright. 
The IR emission of the SNR is well correlated with the X-ray emission, 
being bright in the east and in the south, whereas there seems to be 
little emission from the SNR to the 
northwest.

Emission from the SNR can be seen neither at shorter wavelengths 
(IRAC 3.6, 4.5, and 5.8~$\mu$m) nor at longer wavelengths (\herschel\ 
PACS 160~$\mu$m
and SPIRE 250~$\mu$m, 350~$\mu$m, and 500~$\mu$m). 
In all \spitzer\ IRAC images at $\leq$8~$\mu$m two dark absorption features
can be identified against the bright background at the eastern rim of the SNR 
shell and northeast of the center of the SNR (see Fig.\,\ref{images2}, top 
right).
These are most likely caused by dense foreground clouds.
At 160~$\mu$m, 250~$\mu$m, and 350~$\mu$m some structures are seen in the SNR 
region but these are most likely not associated to the SNR itself.  
A bright spot is seen at longer wavelengths (Fig.\,\ref{images2}, bottom), 
which seems to be associated with a foreground cloud.

\section{Spectral analysis of X-ray data}\label{xspectra}

\begin{table*}
\begin{center}
\caption{\label{spectab}
Best-fit parameters of the analysis of \chandra\ spectra. For the ellipse
region in the bright shell see Table \ref{elltab}.
}
\vspace{-3mm}
\small
\begin{tabular}{lrrrrrrrrrrrr} 
\hline\hline 
Region & \nh & $kT$  & Mg$^\ast$ & Si$^\ast$ & S$^\ast$ & Ar$^\ast$ & Ca$^\ast$ & Fe$^\ast$ & $\tau^\triangledown$ & $norm$ & $\chi^2$ & d.o.f \\
& [cm$^{-2}$] & [keV] &(solar) &(solar) &(solar) &(solar) &(solar) &(solar) & [s\,cm$^{-3}$] & [cm$^{-5}$] & & \\
\hline\noalign{\smallskip}
\multicolumn{13}{c}{VNEI} \\
\hline\noalign{\smallskip}
SNR$^\star$& 3.5$^{+0.2}_{-0.2}$e+22 & 1.1$^{+0.1}_{-0.1}$ & 0.7$^{+0.3}_{-0.1}$ & 0.9$^{+0.1}_{-0.1}$ & 1.3$^{+0.2}_{-0.1}$ & 1.3$^{+0.2}_{-0.2}$ & 1.6$^{+0.6}_{-0.7}$ & 0.3$^{+0.3}_{-0.3}$ & 5.6$^{+0.7}_{-0.4}$e+10 & 4.2$^{+0.4}_{-0.6}$e--2 & 1.21 & 311 \\
\noalign{\smallskip}
1          & 3.1$^{+0.2}_{-0.3}$e+22 & 1.0$^{+0.2}_{-0.2}$ & - & - & - & - & - & - & 6.0$^{+5.1}_{-2.6}$e+10 & 2.1$^{+0.9}_{-0.7}$e--3 & 0.87 & 182 \\
\noalign{\smallskip}
2          & 3.2$^{+0.4}_{-0.3}$e+22 & 1.0$^{+0.3}_{-0.3}$ &  - & - & - & - & - & - & 7.5$^{+15.2}_{-3.3}$e+10 & 2.6$^{+0.9}_{-0.7}$e--3 & 0.97 & 227 \\
\noalign{\smallskip}
3          & 2.1$^{+0.6}_{-0.5}$e+22 & 1.4$^{+0.5}_{-0.3}$ &  - & - & - & - & - & - & 3.7$^{+3.1}_{-1.4}$e+10 & 5.0$^{+3.5}_{-1.8}$e--4 & 1.13 & 83 \\
\noalign{\smallskip}
4          & 3.7$^{+0.1}_{-0.2}$e+22 & 1.0$^{+0.1}_{-0.1}$ &  - & - & 1.3$^{+0.1}_{-0.2}$ & - & - & - & 9.0$^{+2.6}_{-2.0}$e+10 & 1.1$^{+0.2}_{-0.2}$e--2 & 1 17 & 251 \\
\noalign{\smallskip}
5 & 4.1$^{+0.1}_{-0.1}$e+22 & 1.1$^{+0.1}_{-0.1}$ &  - & 1.2$^{+0.1}_{-0.1}$ & 1.6$^{+0.1}_{-0.1}$ & 1.6$^{+0.3}_{-0.2}$  & - & - & 6.4$^{+0.5}_{-0.7}$e+10 & 2.6$^{+0.3}_{-0.3}$e--2 & 1.17 & 233 \\
\noalign{\smallskip}
6          & 3.3$^{+0.2}_{-0.2}$e+22 & 1.1$^{+0.1}_{-0.2}$ &  - & - & 1.5$^{+0.3}_{-0.2}$ & 2.0$^{+0.8}_{-0.8}$ & - & - & 7.8$^{+3.4}_{-1.9}$e+10 & 2.9$^{+0.7}_{-0.5}$e--3 & 0.91 & 157 \\
\noalign{\smallskip}
hard       & 4.4$^{+0.4}_{-0.4}$e+22 & 1.4$^{+0.4}_{-0.3}$ & 1.7$^{+1.3}_{-0.9}$   & 1.7$^{+0.5}_{-0.4}$  & 1.6$^{+0.5}_{-0.4}$ & - & - & - & 6.1$^{+2.4}_{-1.5}$e+10 & 1.1$^{+0.6}_{-0.3}$e--3 & 1.01 & 100 \\
\noalign{\smallskip}
\hline\noalign{\smallskip}
\multicolumn{13}{c}{VPSHOCK} \\
\hline\noalign{\smallskip}
SNR$^\star$& 3.5$^{+0.2}_{-0.3}$e+22 & 1.1$^{+0.1}_{-0.1}$ & 0.8$^{+0.2}_{-0.2}$ & 0.9$^{+0.1}_{-0.1}$ & 1.6$^{+0.1}_{-0.1}$ & 1.5$^{+0.3}_{-0.1}$ & 1.7$^{+0.6}_{-0.7}$ & 0.3$^{+0.4}_{-0.2}$ & 1.1$^{+0.1}_{-0.2}$e+11 & 4.3$^{+0.5}_{-0.6}$e--2 & 1.25 & 311 \\
5 & 4.1$^{+0.1}_{-0.1}$e+22 & 1.1$^{+0.1}_{-0.1}$ &  - & 1.2$^{+0.1}_{-0.1}$ & 1.8$^{+0.1}_{-0.1}$ & 1.9$^{+0.4}_{-0.3}$  & - & - & 1.1$^{+0.2}_{-0.2}$e+11 & 2.9$^{+0.1}_{-0.2}$e--2 & 1.32 & 232 \\
\hline\hline 
\end{tabular}
\end{center}
\vspace{-3mm}
\raggedright
$^\ast$ For the sub-regions 1 to 6, and for the circular region around the hard emission, 
the abundances were fixed to 1.0
if they were consistent with solar values after they had been freed.\\
$^\triangledown$ Upper limit of the ionisation timescale for the VPSHOCK model.\\
$^\star$ Entire SNR.
\end{table*}

For the analysis of the spectrum taken in X-rays, we extracted spectra from 
the newest \chandra\ data with the longest exposure time (ObsID 16766).
Point sources have been removed from the data. 
We have tried different 
extraction regions based on X-ray and IR emission and have defined
areas, which show similar spectral properties in X-rays based on preliminary 
spectral analyses.
By this method we make sure that regions that have similar spectral properties
in X-rays and coinciding brightness distribution between X-rays and IR are
selected. At the same time the size of the regions should be not too small
to have high enough statistics.
The final extraction regions which have been used for the spectral analysis
reported here are shown in Fig.\,\ref{images1}.
We analysed the spectra using the 
X-ray spectral fitting package XSPEC Ver.\ 12.8.2.
All regions were analysed using the model VNEI, which reproduces the
emission of  a non-equilibrium ionisation (NEI) collisional plasma
with variable element abundances.
The spectrum of the entire SNR (in circular region shown in
Fig.\,\ref{images1}, left) has also been studied using the model
VPSHOCK (emission from a plane-parallel shock with a range of ionisation
timescale).
The parameters of the best fits are listed in Table \ref{spectab}.

\begin{figure}
\centering
\includegraphics[trim=40 20 50 40,width=0.49\textwidth]{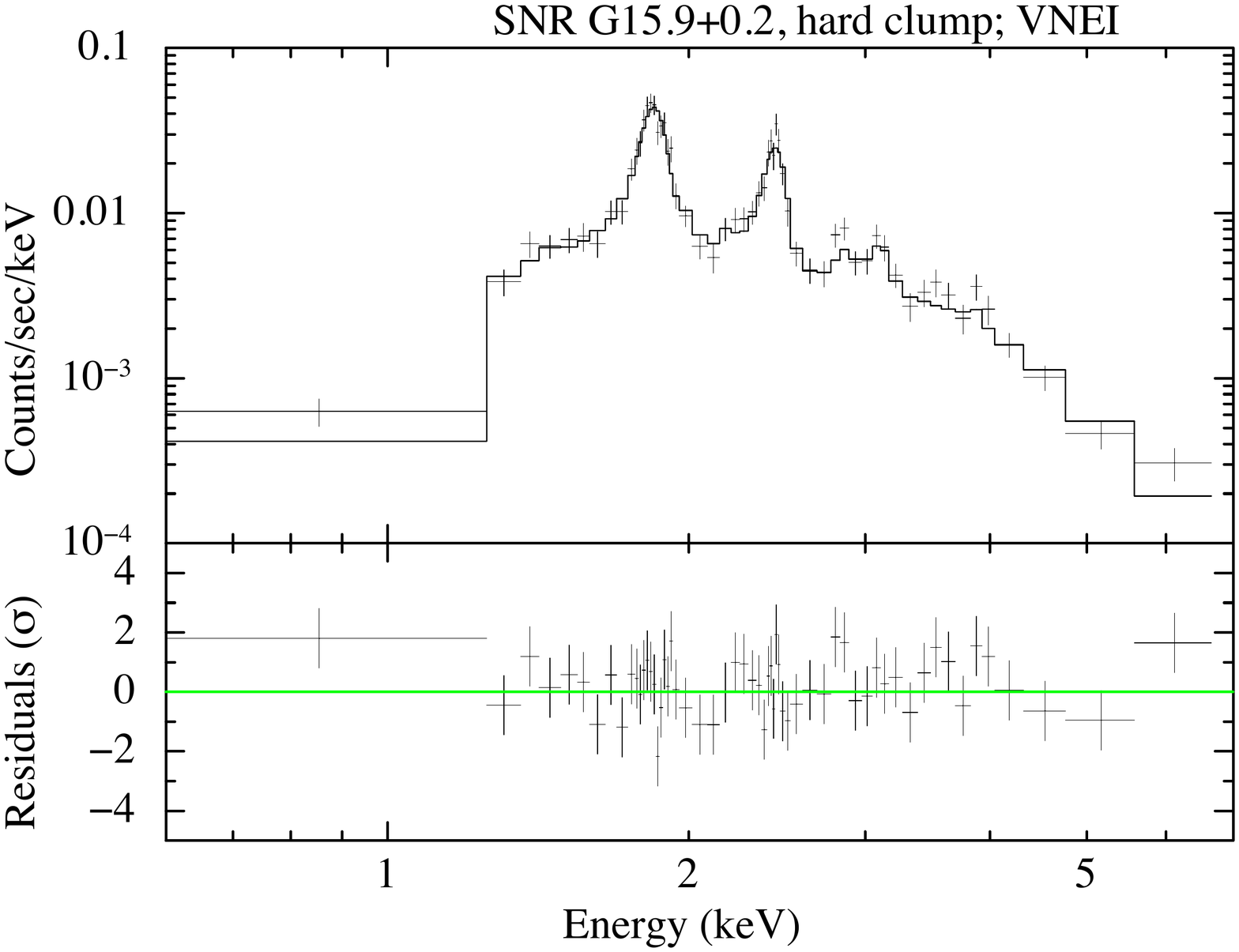}
\includegraphics[trim=40 20 50 40,width=0.49\textwidth]{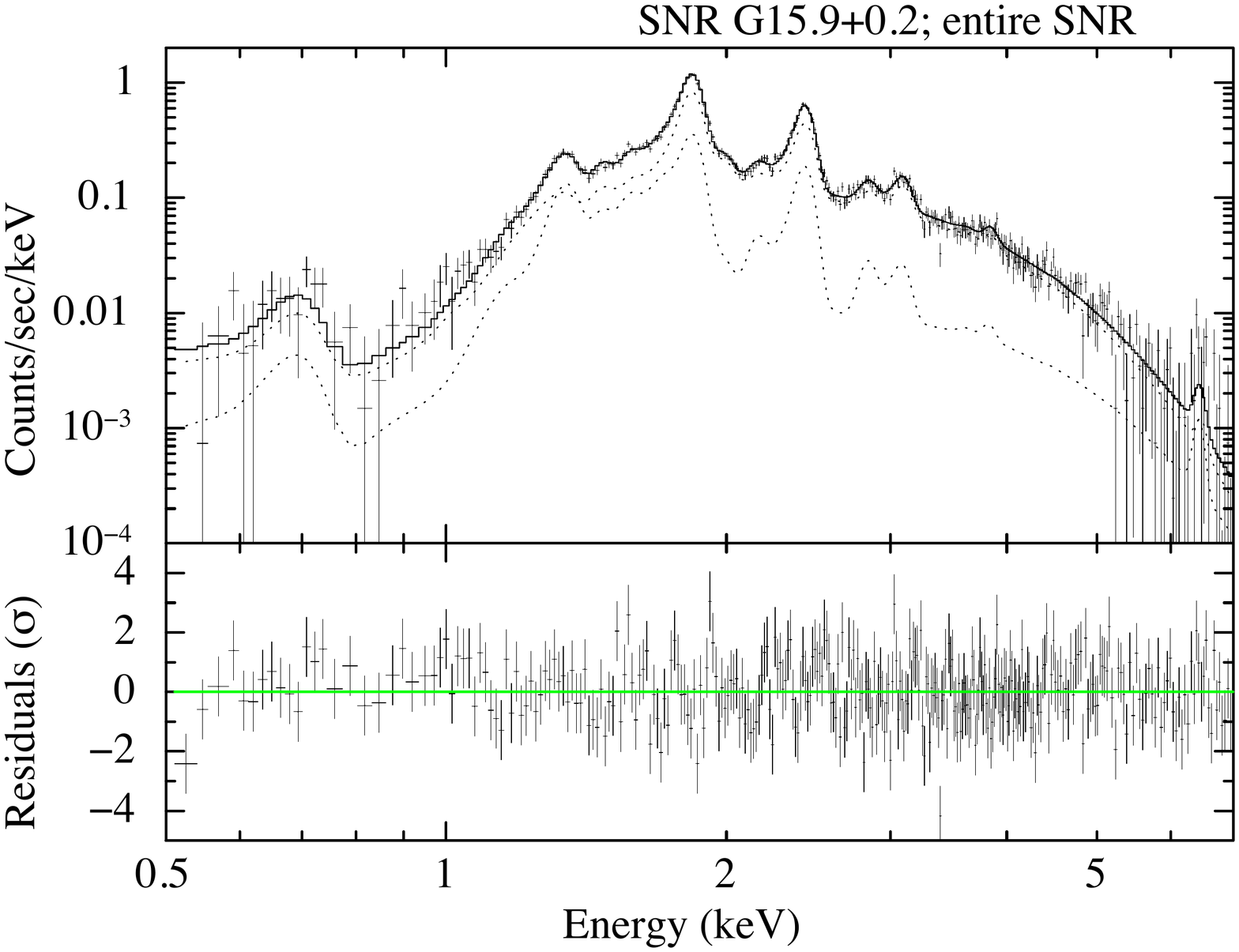}
\includegraphics[trim=40 20 50 40,width=0.49\textwidth]{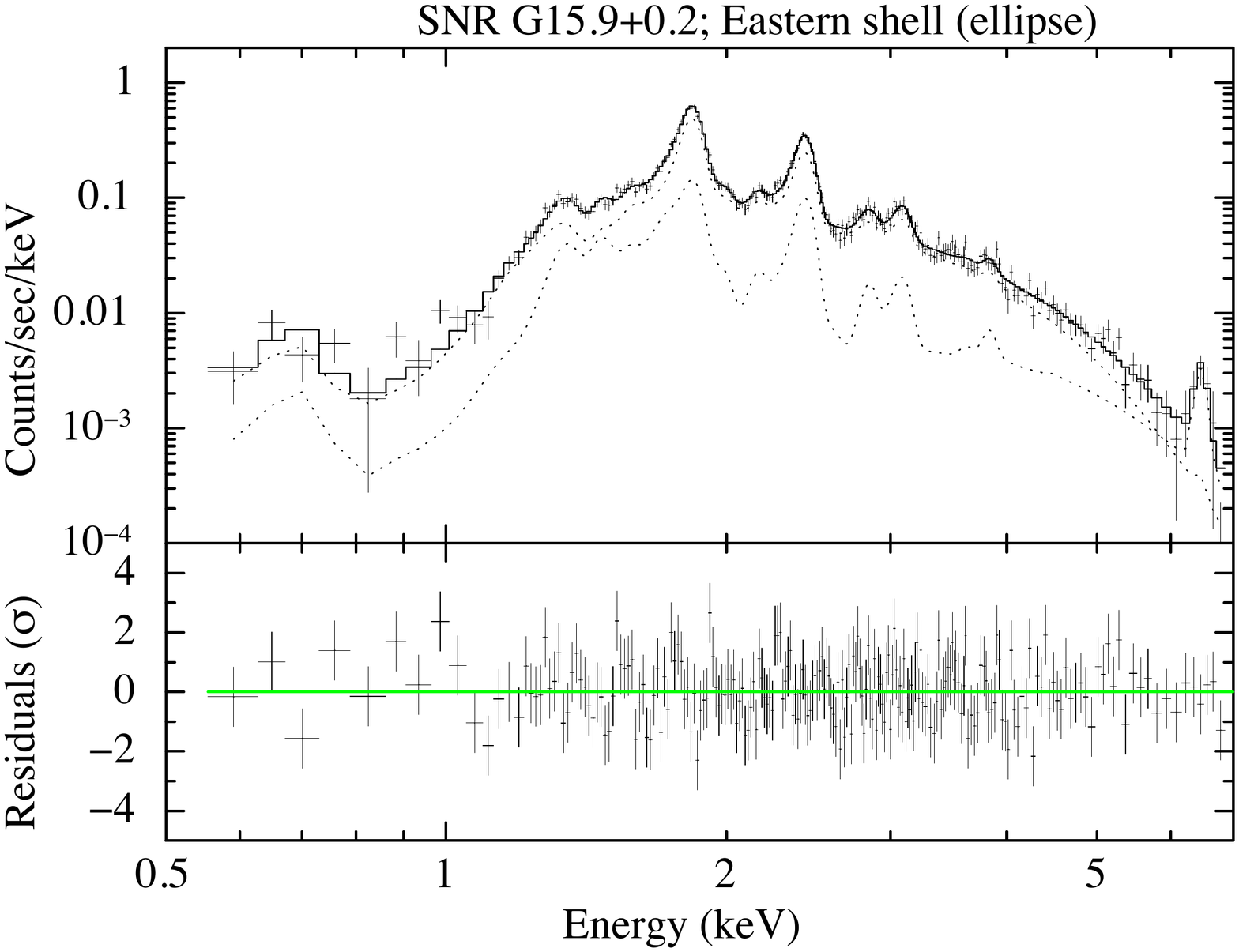}
\caption{\chandra\ ACIS-S3 (ObsID: 16766) spectrum of 
the small blue-ish (`hard') region with the VNEI model (top),
the entire \snr\ with the best fit model (see Sect.\,\ref{specejecta}, middle),
and the ellipse region in the bright eastern shell (bottom).
The fitted models and the residuals are also shown.
\label{spectra}}
\end{figure}

The fits of the entire SNR with VNEI and VPSHOCK models yield parameter
values that are consistent with each other (Table \ref{spectab}, first and 
last rows). 
The average foreground column 
density is \nh\ = 3.5 $\times 10^{22}$ cm$^{-2}$ and the plasma temperature
is $kT$ = 1.1 keV. 
Most of the elements have solar abundances, only
S and Ar yield significantly higher values. 
The ionisation timescale is of the order
of $\tau = nt =  10^{11}$ s cm$^{-3}$, suggesting NEI. 
The absorbed X-ray flux of the SNR is
$F_{\rm abs} (0.3 - 10.0 \mathrm{~keV}) = 5.76 \times 10^{-12}$ erg cm$^{-2}$ s$^{-1}$,
whereas the unabsorbed flux is
$F_{\rm unabs} (0.3 - 10.0 \mathrm{~keV}) = 3.15 \times 10^{-10}$ erg cm$^{-2}$ s$^{-1}$.
The spectrum is shown in Fig.\,\ref{spectra} (middle).

Due to lower statistics, the regions 1 - 6 have only been analysed using the
VNEI model. The spectra of all six regions can be well fitted with a temperature
that is consistent for all regions (and for the fit of the entire SNR). 
Also the ionisation timescale is consistent with each other. Region 3, 
which is the interior region showing rather soft emission 
(appearing red in Fig.\,\ref{images1}, left), has lower foreground \nh\ 
than the other regions. 
Significantly higher \nh\ is obtained for the eastern shell
(region 5),
indicating that there is material on the line of sight. This is 
consistent with the distribution of H$\alpha$ emission, which shows 
absorption in the same region  (Fig.\,\ref{images2}, top left). 
In particular, there seems to be a
dense cloud in front of the SNR at its eastern rim, seen as a dark
absorption feature in the 8.0$\mu$ image (Fig.\,\ref{images2}, top right).
All regions can be nicely fitted with solar abundances for
all elements. Only in regions 4, 5, and 6, freeing the parameter for S 
abundance and Ar abundance in regions 5 and 6 slightly improves the fits. 

The results of the spectral analysis so far show that the X-ray emission of
the SNR is dominated by that of shocked ISM and there is no significant
spectral variation inside the SNR.
However,
there is a small region in the southeastern shell indicated with a yellow 
circle in the X-ray three-colour-image (Fig.\,\ref{images1}, right), 
in which both the X-ray emission and 
the IR emission shows an interior arc-like structure.
Moreover, this region appears blue in the X-ray 
three-colour image and thus seems to show harder X-ray emission. Therefore, we 
also extracted spectrum from this particular region in order to find out what 
the origin of the bright hard X-ray emission might be.
The spectrum can be fitted well with a VNEI model with free abundances for Mg, 
Si, and S (see Table \ref{spectab}). 
These three elements turn out to be slightly overabundant. If one looks at the spectrum
there is also a hint of excess at the position of the Fe K line, however, the 
statistics are too low to yield any significant fit parameter for the Fe 
abundance.
Since this X-ray bright region is not bright in IR and the spectrum indicates
overabundance of $\alpha$ elements, there might be a clump of heated ejecta 
at this position. 
The high foreground absorption does not allow us to determine abundances 
for O or Ne.
We also looked for a non-thermal component in the spectrum. 
Adding a power-law component did not improve the fit but resulted in rather 
unrealistic values for the power-law index. Therefore, we conclude that the
emission is fully thermal.

\subsection{Ejecta emission}\label{specejecta}

Our spectral parameters \nh, $kT$, and $\tau$ are consistent with
the results obtained by \citet{2017A&A...597A..65M}. They also
find a slight overabundance of elements Mg, Si, S, Ar, and Ca in various
regions of the SNR. In addition, an excess at the position of the
Fe K line at $\sim6.5$ keV was found in the spectrum of the eastern part of the
SNR shell. The \chandra\ spectra 
of the regions studied so far
are all consistent with ISM abundance for
Fe or indicate a lower Fe abundance. It is difficult to constrain
the Fe abundance parameter since the spectrum is heavily absorbed below 1 keV
where the Fe L emission would be observable.

In order to search for ejecta emission discovered by
\citet{2017A&A...597A..65M} in the \xmm\ spectrum, we extracted the spectrum
of a larger region in the brighter eastern shell as indicated by the
ellipse in Fig.\,\ref{images1} (left).
As can be seen in the spectrum shown in Fig.\,\ref{spectra} (bottom),
there is a clear feature at 6.5 keV which is likely Fe K line emission.
While this line was not significant in the spectrum of the entire SNR
(see Fig.\,\ref{spectra}, middle), which also includes large portions with
very faint emission, the source emission is bright enough with respect
to the local X-ray background in the ellipse region to make the Fe K line
detectable.
We thus confirm the emission detected by \citet{2017A&A...597A..65M}
and are able to study the ejecta emission using the spectrum of the 
bright eastern shell.

\begin{table}
\begin{center}
\caption{\label{elltab}
Spectral fit results of the ellipse region in the bright shell.
}
\vspace{-3mm}
\small
\begin{tabular}{lrr} 
\hline\hline 
Parameters & Model 1 & Model 2 \\
\hline\noalign{\smallskip}
\nh\ [cm$^{-2}$] & 3.9$^{+0.1}_{-0.1}$e+22 & 3.8$^{+0.1}_{-0.1}$e+22 \\
\noalign{\medskip}
$kT_{\rm NEI}$ [keV] & 1.0$^{+0.1}_{-0.2}$ & 0.9$^{+0.1}_{-0.1}$ \\
$\tau_{\rm NEI}$  [s\,cm$^{-3}$] & 7.6$^{+0.6}_{-0.6}$e+10 & 1.0$^{+0.6}_{-0.3}$e+11\\
$norm_{\rm NEI}$  [cm$^{-5}$] & 3.2$^{+0.3}_{-0.3}$e--2 & 3.9$^{+0.4}_{-0.3}$e--2 \\
\noalign{\medskip}
$kT_{\rm VNEI}$ [keV] & & 3.0$^{+0.7}_{-0.7}$ \\
Mg$_{\rm VNEI}$ & & 34$^\dagger$ \\
Si$_{\rm VNEI}$ & & 8$^\dagger$ \\
S$_{\rm VNEI}$ & & 9$^\dagger$ \\
Ar$_{\rm VNEI}$ & & 5$^\dagger$ \\
Ca$_{\rm VNEI}$ & & 5$^\dagger$ \\
Fe$_{\rm VNEI}$ & & 14$^\dagger$ \\
$\tau_{\rm VNEI}$  [s\,cm$^{-3}$]& & 2.6$^{+0.9}_{-0.5}$e+10 \\
$norm_{\rm VNEI}$ [cm$^{-5}$] & & 6.3$^{+2.3}_{-2.4}$e--4 \\
\hline
$\chi^2$ & 1.8 & 1.0 \\
d.o.f & 220 & 216 \\
\hline\hline 
\end{tabular}
\end{center}
\vspace{-3mm}
\raggedright
$^\dagger$ Not well constrained (see text).
\end{table}

We first fitted the spectrum only assuming one NEI component with solar 
abundances. The best-fit parameters (see Table \ref{elltab}, Model 1) are
similar to that of other shell regions (Table \ref{spectab}), but the
reduced $\chi^2$ = 1.8 (with degrees of freedom d.o.f\ = 220) is high.
We have therefore included an additional thermal component to model the
ejecta emission, while the first component is used to model the ISM emission.
For this additional ejecta component, we first set all element abundances 
to zero and fit the abundances of the elements Mg, Si, S, Ar, Ca, and Fe by
freeing the parameters one by one. The abundances of these elements become
significantly higher than solar. 
The fit results of the initial NEI component with solar abundances for all 
elements, are  $kT = 0.89~(+0.03, -0.05)$ keV and 
$\tau = 1.0~(+0.6, -0.3) \times 10^{11}$ s cm$^{-3}$.
The best fit has a reduced $\chi^2$ of 1.0 for 216 degrees of freedom
(see Table \ref{elltab}, Model 2). 
Since the normalisation and the element abundances are strongly dependent on
each other, the parameters of the second component are not well constrained. 
The fitted values relative to solar abundances are
Mg:Si:S:Ar:Ca:Fe = 34:8:9:5:5:14.
The best fit absorbing column density for the whole model is 
\nh\ = $3.8~(+0.1, -0.1) \times 10^{22}$ cm$^{-2}$.
This yields an absorbed X-ray flux for the ISM 
component of the eastern (E) shell of
$F_{\rm abs, ISM, E} (0.3 - 10.0 \mathrm{~keV}) = 4.5 \times 10^{-12}$ erg cm$^{-2}$ s$^{-1}$,
and for the ejecta component
$F_{\rm abs, ejecta, E} (0.3 - 10.0 \mathrm{~keV}) = 1.3 \times 10^{-12}$ erg cm$^{-2}$ s$^{-1}$.

As shown before, the ejecta component is not detected significantly in the
spectrum of the entire SNR. However, in order to get an estimate of the
emission components in the entire SNR, we freeze all parameters of the two
additive
spectral components except for the normalisations and use the model for
the fit of the spectrum of the SNR. We thus keep the shape of the spectral
components but rescale their fluxes. This method yields a spectral fit
with a reduced $\chi^2$ of 1.1 for 315 degrees of freedom.
For the SNR, we obtain an absorbed X-ray flux for the ISM 
component of 
$F_{\rm abs, ISM} (0.3 - 10.0 \mathrm{~keV}) = 4.3 \times 10^{-12}$ erg cm$^{-2}$ s$^{-1}$,
and for the ejecta component
$F_{\rm abs, ejecta} (0.3 - 10.0 \mathrm{~keV}) = 1.6 \times 10^{-12}$ erg cm$^{-2}$ s$^{-1}$.
The spectra with the best-fit models are shown in Fig.\,\ref{spectra} (middle
and bottom).

\section{Infrared fluxes}\label{irflux}

To study the IR emission from the SNR, we measured the SNR flux densities in an aperture 
with a radius of 160\arcsec\ and
estimated the local background 
in a region located south of the SNR.
The background region  was selected to have low IR surface 
brightness and exclude any prominent IR features. 
The other directions outside the SNR contain external emission in the southeast 
direction, SNR dust in the north and an apparent brighter edge of a cloud 
in the northeast direction (``NE rim''). Therefore, we did not use an
annulus around the source aperture to estimate the background.
Point sources were masked in the IRAC 8.0 and 24~$\mu$m images. 
The \herschel\ images do not contain point sources. The dark absorption cores 
seen at 8.0 and 24~$\mu$m are masked out, 
and their emission counterparts in the \herschel\ images are also masked (these 
features include the foreground clump in the northwest 
of the SNR center and the eastern rim 
clumps) if they are seen in the images. Furthermore, 
at 8~$\mu$m image artefacts ("jailbars") were masked. 
The median value of the sky brightness was used as the background value.
The background-subtracted aperture flux density was then converted into Janskys.

The measured flux densities were aperture-corrected, but not colour- or 
extinction-corrected. Omitting the colour-correction will produce typically an 
error from a few to 10~\% percent. 
The flux density values for the SNR are $3.2 \pm 1.5$, $15.9 \pm 4.0$, and  
$88.5 \pm 9.5$~Jy for 8.0, 24, and 70~$\mu$m, respectively.
The \snr\ IR flux densities have been previously computed as a part of 
larger SNR surveys. \citet{pinheirogoncalvesetal11} obtained aperture-corrected 
values at 8.0, 24, and 70~$\mu$m of $3.2 \pm 2.5$, $16 \pm 2$, and $55 \pm 16$~Jy, 
respectively, using \spitzer\ data. \citet{2016ApJ...821...20K} 
used \spitzer\ MIPS 
24~$\mu$m and \herschel\ PACS 70~$\mu$m observations to derive fluxes of 
$10.1 \pm 0.4$ and $49.1 \pm 2.7$~Jy, respectively.

In addition to the SNR, we had 
inspected the exterior NE rim properties to verify whether
the region could be used for extracting the background value for the SNR. 
We compared the surface brightnesses in the NE rim and the chosen 
background region. For this no background value was subtracted of either 
region. The results are listed in Table \ref{ir:sbcomparison}. At all 
observed wavelengths the surface brightness in the background region 
is lower. Comparing the IR images with the H$\alpha$ image suggests 
that the NE rim is the brightened edge of the large foreground cloud 
seen in H$\alpha$. This together with the surface brightness values 
suggests that the NE rim is an object separate from the SNR. The derived 
background-subtracted flux densities for NE rim are $1.7 \pm 1.3$, 
$1.2 \pm 1.1$, $15.1 \pm 3.9$, and $51.3 \pm 7.2$~Jy, at 8, 24, 70, 
and 160~$\mu$m, respectively. These are not aperture-corrected, and
the corrections introduce at the most a $\sim$10\% error.

\begin{table}
\centering
\caption{Comparison of surface brightness in the NE rim and the background
regions.}
\begin{tabular}{rrr}
\hline\hline
Wavelength & Background & NE rim \\
$[\mu\mathrm{m}]$ & [MJy/sr]   & [MJy/sr] \\
\hline
8	& 59.7  & 69.8 \\
24	& 70.3  & 77.4 \\
70	& 7.0   & 93.7 \\
160	& 409.2 & 700.9 \\
\hline \hline
\end{tabular}
\label{ir:sbcomparison}
\end{table}

\section{Discussion}

\subsection{SNR properties assuming Sedov solution}

The morphology and the X-ray spectrum of \snr, which is consistent with 
ISM emission, indicate that it is in the Sedov phase. 
Therefore, we calculate its physical 
parameters using the Sedov-Taylor-von Neumann similarity solution 
\citep{1959book..........S,1950PRSLA..201..159T,1947LASLTS.......vN}.
For cosmic abundances \citep{1989GeCoA..53..197A}, the gas density is 
$n_{\rm gas} = 1.1~n_{\rm H}$, where $n_{\rm H}$ is the atomic H number 
density. For fully ionised plasma which can be expected for an X-ray SNR, 
the electron number density is for cosmic abundances 
is $n_{\rm e} = 1.21~n_{\rm H}$.
Therefore, the total number density of the gas is
$n = n_{\rm e} + n_{\rm i} = 2.3~n_{\rm H}$ with a mean
mass per free particle $\bar{m} = 0.61~m_{\rm p}$, 
with proton mass $m_{\rm p}$.
We estimate the shock velocity using the mean shock temperature 
$kT$ = 0.89 $\pm$ 0.05 keV for the ISM component obtained from the fit 
of the ellipse region in the eastern shell as well as the entire SNR 
assuming ISM and ejecta components.
With
\begin{equation}\label{tempvel}
v_{\rm s} = \left(\frac{16 k T_{\rm s}}{3 \bar{m}}\right)^{1/2}, 
\end{equation}
where $k$
is the Boltzmann's constant,
the shock velocity is $v_{\rm s}$
= 860 $\pm$ 30 km~s$^{-1}$.

If we ignore the northern blow-out, the SNR can be regarded as a nicely 
circular shell with a radius of 
$\theta$ = 155\arcsec $\pm$ 5\arcsec = 
$(7.5 \pm 0.3) \times 10^{-4}$ rad as can be verified in the deep \chandra\ 
mosaic image (see Fig.\,\ref{images1}, right). 
The linear radius can be written as 
$R_{\rm s} = (7.5 \pm 0.3) \times d_{\rm 10 kpc}$ pc 
or
$R_{\rm s} = (2.3 \pm 0.1) \times 10^{19} \times d_{\rm 10 kpc}$ cm
with $d_{\rm 10 kpc}$ being the distance normalised to 10 kpc.
Unfortunately, the distance $d$ to \snr\ is not yet known. 
\citet{1982MNRAS.200.1143C} obtained a distance of 16.7 kpc based on a
$\Sigma$-$D$-relation, which would 
result in a rather unrealistic 
large size with a diameter of 27 pc. High foreground absorption indicates
that the SNR is most likely located close to or on the other side of
the Galactic center. Therefore, \citet{2006ApJ...652L..45R} used a fiducial 
distance of 8.5~kpc in their calculations.
We thus have $d_{\rm 10 kpc} = 0.85 - 1.67$.
The age of the remnant can be estimated from the shock velocity using the
Sedov similarity solution:
\begin{equation}\label{simvel}
t = \frac{2 R_{\rm s}}{5 v_{\rm s}}
\end{equation}
yielding an age estimate of $t$
= $(1.1 \pm 0.1) \times 10^{11} d_{\rm 10 kpc}$ s
= $(3400 \pm 200)  \times d_{\rm 10 kpc}$ yr.

From the normalisation of the NEI or the VNEI, which is defined as
$norm = 10^{-14}/(4 \pi d^{2}) \times \int n_{\rm e} n_{\rm H} dV$
[cm$^{-5}$],
one can estimate the densities.
We rearrange the integral
$\int n_{\rm e} n_{\rm H} dV$ using $n_{\rm e} = (n_{\rm e}/n_{\rm H}) n_{\rm H}$
and $n_{\rm e}/n_{\rm H} = 1.21 = const$ into
$(n_{\rm e}/n_{\rm H}) \int n_{\rm H}^{2} dV 
=(n_{\rm e}/n_{\rm H}) \int_{0}^{R_{\rm s}} 4 \pi R^2 n_{\rm H}^{2} dR
=(n_{\rm e}/n_{\rm H}) 2.06 n_{\rm H,0}^{2} (4/3) \pi R_{\rm s}^3$.
For the integration in the last step, the calculation by 
\citet{1975ICRC...11.3566K} for the Sedov solution has been used
\citep[see, e.g.,][]{1982ApJ...253..268C}.
Therefore, we get
$norm$ 
$= 8.3 \times 10^{-15} n_{\rm H,0}^{2} R_{\rm s}^3 d^{-2}$.
Using the normalisation obtained for the ISM component of the spectrum
of the entire SNR,
the initial H density is
$n_{\rm H,0} = 3.0 \times (norm/d_{\rm 10 kpc})^{1/2}
= (0.73 \pm 0.06) \times d_{\rm 10 kpc}^{-1/2}$ cm$^{-3}$
and the gas density is 
$n_{\rm gas,0} = 1.1 n_{\rm H,0} = (0.81 \pm 0.06) 
\times d_{\rm 10 kpc}^{-1/2}$ cm$^{-3}$.
If we assume that this is the mean number density of the interstellar gas
in which the SNR is expanding, the mass of ISM that the SNR has shocked is
$M_{\rm ISM} = 4\pi/3 R_{\rm s}^3 \times 1.4 n_{\rm H,0} m_{\rm p}
= 44 M_{\sun} \times d_{\rm 10 kpc}^{2.5}$.

For a distance in the range of $d$ = 8.5 -- 16.7 kpc, 
the SNR age and the ambient H density are
$t$ = (2900 $\pm$ 200) -- (5700 $\pm$ 300) yr and 
$n_{\rm H,0}$ = (0.79 $\pm$ 0.07) -- (0.56 $\pm$ 0.05) cm$^{-3}$, 
respectively. 
These numbers are realistic for both extreme values of the distance and
do not allow us to constrain the distance.

The circular morphology of the SNR and the distribution of the cold ISM as 
seen in
the IR images suggest that the SNR is expanding in a region with a higher
density in the south and the east and lower density in particular in the 
northwest. Therefore, the outer shock could expand further in the northwest
than in the rest of the SNR. 
On this lower-density side of the ISM, the evolution of the SNR is not 
significantly affected by the existence of higher-density gas on the other side.
Therefore, one can apply the Sedov solution also for this part of the SNR. 
The distance of the outer rim in the north from
the geometrical center of the circle around the bright main part of \snr\
is $\theta_{\mathrm n}$ = 210\arcsec $\pm$ 10\arcsec, corresponding to
$R_{\rm s,n} = (10.2 \pm 0.5) \times d_{\rm 10 kpc}$ pc 
$= (3.1 \pm 0.1) \times 10^{19} \times d_{\rm 10 kpc}$ cm.
If we now use the age which was derived from the rest of the remnant
we can calculate the expansion velocity and the ambient density based on
the Sedov solution: 
$v_{\rm s,n} = 2 R_{\rm s,n}/(5 t)$
= $1100 \pm 100$ km~s$^{-1}$.
Since for the Sedov solution, we have the relation:
$R_{\rm s} \propto n_{\rm gas,0}^{-1/5} t^{2/5}$, 
and the age $t$ is obviously the same for the northern part and for the rest 
of the SNR, one gets:
$n_{\rm gas,0,n} = R_{\rm s}^5 /  R_{\rm s,n}^5 n_{\rm gas,0}$
= $(0.18 \pm 0.02) \times d_{\rm 10 kpc}^{-1/2}$ cm$^{-3}$.
Therefore, there is a change in ambient density of a factor of $\sim$4.6
in $\sim$20 pc, which allowed the shock to propagate faster 
in the north. 
In principle, this difference in velocity should also be observable in the
spectrum of the thermal X-ray emission, which should have a higher 
temperature. However, the very low statistics in the northern blow-out
region makes it not possible to measure the temperature of the
emitting plasma.
On the other hand, the interaction of the SNR with the denser ISM in the
south and in the east resulted in higher emissivity and thus brighter
X-ray emission. Furthermore, the shocks also caused the heating of dust
on this side,
which makes the SNR detectable also in IR at 24 $\mu$m and 70 $\mu$m.

So far, not many SNRs with CCOs have been found.  
Since the morphology and the evolution of an SNR strongly depend on the 
geometry of the progenitor and
the  circumstellar/interstellar environments in which it is located,
the study of SNRs are crucial for a better understanding of the formation 
mechanisms of different types of neutron stars.
The most famous SNR with a CCO with an age which is about one order of
magnitude lower is SNR Cas A.
For Cas A an age of $\sim$320 yr
\citep{1980JHA....11....1A,2006ApJ...645..283F} and
a shock velocity of 5200 km s$^{-1}$ was measured.
A major difference between \snr\ and Cas A is that the reverse shock is
still in the process of heating the ejecta and is observable in Cas A 
\citep{2001ApJ...552L..39G,2004ApJ...614..727M} 
and that its outer shock has traveled through
the circumstellar medium not too long ago, while
\snr\  is well in the Sedov phase. 
\citet{2014ApJ...786...55A} derived a mass of 0.04$M_{\sun}$ for the
shocked hot component of the dust inside Cas A, while the still
unaffected cold dust component in the unshocked ejecta is estimated to 
be $\la$0.1$M_{\sun}$.

\subsection{Ejecta mass}

In Section \ref{specejecta} we showed that the X-ray spectrum of the entire SNR can be fit well with
a model consisting of a solar-abundance ISM component and an ejecta component with higher abundances
for elements Mg, Si, S, Ar, Ca, and Fe. 
Assuming a homogeneous density for ejecta, we can estimate the ejecta mass from the spectral parameters.
We use the the normalisation parameter of the ejecta component of the
spectrum of the entire SNR, which is 
$norm = 2.1 (-0.6, +1.6) \times 10^{-3}$ cm$^{-5}$.
Using the assumed solar abundance values for the spectral fit
\citep{1989GeCoA..53..197A} and the fitted abundance values relative to solar,
we obtain
$M_\text{Mg}$ = 0.0536 $\times d_{\rm 10 kpc}^2 M_{\sun}$,
$M_\text{Si}$ = 0.0139 $\times d_{\rm 10 kpc}^2 M_{\sun}$,
$M_\text{S}$ = 0.00828 $\times d_{\rm 10 kpc}^2 M_{\sun}$,
$M_\text{Ar}$ = 0.00135 $\times d_{\rm 10 kpc}^2 M_{\sun}$,
$M_\text{Ca}$ = 0.000709 $\times d_{\rm 10 kpc}^2 M_{\sun}$, and
$M_\text{Fe}$ = 0.0632 $\times d_{\rm 10 kpc}^2 M_{\sun}$ for the
part of the ejecta, which has been ionised and heated by the reverse shock.
These values have large uncertainties, since the abundance parameters together 
with the $norm$ parameter are not well constrained. The comparison to the 
nucleosynthesis yields for core-collapse SNe presented by
\citet{1997NuPhA.616...79N}
shows that the estimated masses and abundance ratios suggest a progenitor
mass of $\la20~M_{\sun}$.

\subsection{Dust properties}

We derive the dust temperature $T_{\rm d}$ from the IR flux ratios. 
We assume that the emission comes mainly from shocked hot dust that can be 
described with a single temperature $T_{\rm d}$ and a modified black-body 
emission: 
\begin{equation}
F_{\nu} \propto \nu^{\beta} B_{\nu}(T_{\rm d})
\end{equation}
where $F_{\nu}$ is the flux density, $\beta$ is the dust emissivity index which depends 
on the dust composition ($\beta\sim1-2$, we use $\beta=2$), and 
$B_{\nu}(T_{\rm d})$ is the Planck function evaluated at $T_{\rm d}$. 
Following this we can solve $T_{\rm d}$ from the ratio 
\begin{equation}
\frac{F_{\nu_{1}}}{F_{\nu_{2}}} = \left(\frac{\nu_{1}}{\nu_{2}}\right)^{\beta} \frac{B_{\nu_{1}}(T_{d})}{B_{\nu_{2}}(T_{\rm d})}.
\end{equation}
Using the measured flux densities for 24 and 70~$\mu$m (Sect.\,\ref{irflux}) 
we get 55.9~K. \citet{pinheirogoncalvesetal11} obtained 60~K, and 
\citet{2016ApJ...821...20K} 57.8~K using the same ratio $F_{24}/F_{70}$.

For the outside NE rim, the flux density ratios $F_{24}/F_{70}$ and $F_{70}/F_{160}$ 
yield temperatures of 50.3~K and 21.6~K, respectively. The flux density at 24~$\mu$m is 
very low and has relatively high errors, but within the $F_{24}$ 1-$\sigma$ error the derived 
temperature is lower than the one derived for the SNR. For the SNR dust we cannot derive a flux 
density value at 160~$\mu$m as there is no discernible emission from the SNR. 
We can however measure the emission at 160~$\mu$m and use the error to estimate an upper limit for the flux density of the SNR dust. The 3-$\sigma$ upper limit for the flux density $F_{160}$ is 35.1~Jy.
Using the $F_{160}$ upper limit, the ratio $F_{70}/F_{160}$ corresponds to a temperature of 37.1~K for the SNR. Flux densities lower than the $F_{160}$ upper limit would result in even higher temperature values.
This suggests that the dust at the NE rim is colder than the SNR dust.

The total IR flux $F_{\rm IR}$ can be estimated using
\begin{equation}
\frac{F_{\rm IR}}{\nu F_{70}} = \frac{\int \kappa_{\nu}B_{\nu}(T_{\rm d}) d\nu}{\nu \kappa_{\nu}B_{\nu}(T_{\rm d})}
\end{equation}
where $\kappa_{\nu}$ is the mass absorption coefficient 
($\kappa_{\nu} \propto \lambda^{-\beta}$). \citet{2016ApJ...821...20K} 
calculated $F_{\rm IR} =  4.32\times10^{-9}$\,erg\,cm$^{-2}$\,s$^{-1}$. 
Our values for $F_{70}$ and $T_{\rm d}$ 
give $F_{\rm IR} = 7.21\times10^{-9}$\,erg\,cm$^{-2}$\,s$^{-1}$.

If the observed IR emission is solely due to thermal dust emission, the
SNR dust mass can be estimated using the following equation:
\begin{equation}
M_{\rm dust}=\frac{d^{2} F_{\nu}}{\kappa_{\nu} B_{\nu}(T_{\rm d})}
\end{equation}
where $d$ is the distance to the SNR, and $\kappa_{\nu}$ the dust mass 
absorption coefficient. We evaluate the dust mass at 24~$\mu$m based on the 
derived temperature 55.9~K. For $\kappa_{\nu}$ we use the dust model by 
\citet{drainearaa03,draineapj103} for a mixture of silicates and carbonaceous 
grains. At 24~$\mu$m the dust mass absorption coefficient is 
505.9~cm$^{2}\,$g$^{-1}$ which 
yields $M_{\rm dust}=0.24 \times d_{\rm 10 kpc}^{2} M_{\odot}
=0.17-0.67 M_{\odot}$
when $d_{\rm 10 kpc}$ = 0.85 -- 1.67.
\citet{pinheirogoncalvesetal11} derived a dust mass of 0.081~$M_{\odot}$ using 
a distance of 8.5~kpc.

We use the \nh\ derived from the X-ray modelling to estimate the 
extinction correction at 24~$\mu$m: $N_{\rm H}=3.5\times10^{22}$~cm$^{-2}$ 
corresponds to $A_{V}=18.7$~mag and hence
$A_{Ks}=2.1$~mag using the extinction law adapted from 
\citet{cardellietal89} and $A_{\rm 24}=0.72$~mag \citep{chapmanetal09}. 
The extinction-corrected flux density is then $1.96 \times F_{24}$, 
leading to a derived 
dust temperature $T_{\rm d}=61.9$~K and a dust mass of 
$0.16 \times d_{\rm 10 kpc}^{2} M_{\odot}$.
Using the shocked ISM mass of 
$M_{\rm ISM} = 44 M_{\sun} \times d_{\rm 10 kpc}^{2.5}$,
the dust-to-gas mass ratio is $M_{\rm dust}/M_{\rm ISM} = 0.0036 \times 
d_{\rm 10 kpc}^{-0.5} = 0.0028 - 0.0039$, a little less but still
comparable to the average dust-to-gas mass ratio measured in the ISM of the 
Milky Way and nearby galaxies \citep{2007ApJ...663..866D}.

However, we note that the use of a single-temperature dust 
population in thermal equilibrium is a highly simplified view of the situation.
Furthermore, even though the SNR is very young and not yet fully in the 
radiative phase, some of the observed IR flux can also be 
from line emission instead of thermal dust emission. 
This causes our dust mass estimate to be an upper limit.

\subsection{IR-X comparison}

The kinetic energy of the SNR shock is dissipated by interaction of charged 
particles and the electromagnetic fields and thus heat the plasma, and 
subsequently, the dust.
The efficiency of converting the kinetic 
energy of the shock into IR radiation of dust $\epsilon_{\rm IR}$
can be estimated using the relation 
\begin{equation}\label{effdef}
\epsilon_{\rm IR} = L_{\rm IR} / \bigg( \frac{1}{2}  n_{\rm H,0} \mu_{\rm H} v_s^3 \times 4 \pi R_s^2 \bigg) 
\end{equation}
with $n_{\rm H,0}$ being the ambient density, 
$\mu_{\rm H} = 1.4$ the mean relative mass per hydrogen atom,
$v_s$ the SNR shock velocity,
$R_s$ the SNR radius
\citep{2016ApJ...821...20K}.
If one compares the X-ray images (Fig.\,\ref{images1}) 
with the IR image  (Fig.\,\ref{images2}) it is striking that
the X-ray emission agrees well with structures in the IR shell.
Using the parameters obtained from the new X-ray and IR analyses, 
we get for the IR conversion efficiency 
$\epsilon_{\rm IR} 
= 0.024 d_{\rm 10 kpc}^{1/2}$.
This value is one to two orders of magnitude lower than the value obtained by
\citet{2016ApJ...821...20K} for well-known mixed-morphology SNRs. 
Compared to these SNRs, the morphology of which is believed to be due to 
shocked emission from shocked interstellar clouds, \snr\ is younger and is 
expanding in a low-density medium.

\section{Summary}

We have studied the X-ray emission of \snr\ using the newest \chandra\ data.
The spectrum of the SNR is consistent with that of thermal emission from 
shocked ISM. In a smaller region in the shell we also found some enhancement
of the silicon and sulfur abundances, which might indicate  emission
from an ejecta clump. 
The spectrum of the bright eastern part of the shell shows a clear Fe K line
at $\sim$6.5 keV. This spectrum has been modelled with two thermal components,
simulating shocked ISM and shocked ejecta. Using the same model components,
we model the spectrum of the entire SNR. Based on a comparison with 
nucleosynthesis yields calculations, we estimate that the progenitor had a
mass of $\la$ 20 $M_{\odot}$.
For the possible range for the distance of $d$ = 8.5 -- 16.7 kpc, the
parameters derived from the analysis of the X-ray emission yield reasonable
values 
for the SNR age of $t$ = (2900 $\pm$ 200) -- (5700 $\pm$ 300) yr and the H 
density of $n_{\rm H,0}$ = (0.79 $\pm$ 0.07) -- (0.56 $\pm$ 0.05) cm$^{-3}$, 
not allowing to constrain the distance.
The total mass of shocked ISM is $M_{\rm ISM}$ = (30 -- 160) $M_{\sun}$.

To study the morphology of the SNR and its relation with the ambient ISM,
we also used H$\alpha$ data from the SuperCOSMOS H$\alpha$ survey and 
infrared data of \herschel\ and  \spitzer. These data at lower energies
indicates that the SNR is located in a region with a complex distribution 
of colder gas and dust.
There is a negative density gradient from southeast to 
northwest, which is responsible for the shape of the SNR
which indicates a blow-out to the north.
We derived a 
dust temperature and mass of
$T_{\rm d}=61.9$~K and SNR dust mass 0.12 -- 0.45~$M_{\odot}$,
corresponding to a dust-to-gas mass ratio of
$M_{\rm dust}/M_{\rm ISM} = 0.0028 - 0.0039$.
Based on IR flux and the shock parameters obtained from the X-ray analysis,
we have shown that the shock-IR-conversion efficiency is low and suggest that, 
even though the morphologies in IR and X-ray match very well, only a small
fraction of the energy has been used for the heating of dust.

\section*{Acknowledgements}

We thank the anonymous referee for the comments that helped
to improve the paper a lot.
M.S.\ acknowledges support by the Deutsche Forschungsgemeinschaft (DFG)
through the Heisenberg fellowship SA 2131/3-1 and the Heisenberg
professor grant SA 2131/5-1.
M.M.\ is funded by the DFG grant SA 2131/4-1.
V.S.\ acknowledges support by the DFG through the grant WE 1312/51-1
and the Russian Government Program of Competitive Growth of Kazan Federal 
University.
\chandra\ X-ray Observatory is operated by the Smithsonian Astrophysical 
Observatory and the National Aeronautics Space Administration under contract 
NAS8-03060.
Herschel is an ESA space observatory with science instruments provided by 
European-led Principal Investigator consortia and with important participation 
from NASA. This work is based in part on observations made with the Spitzer 
Space Telescope, which is operated by the Jet Propulsion Laboratory, California 
Institute of Technology under a contract with NASA.
This research has made use of the NASA/IPAC Infrared Science Archive, which is 
operated by the Jet Propulsion Laboratory, California Institute of Technology, 
under contract with the National Aeronautics and Space Administration, 
the SIMBAD database, and the VizieR catalogue access tool operated at CDS, 
Strasbourg, France.



%
\bibliographystyle{mnras}

\bsp	
\label{lastpage}
\end{document}